\documentclass[aps,prl,twocolumn,nobibnotes,superscriptaddress]{revtex4-2}
\usepackage{inputenc}
\usepackage{textcomp}
\usepackage{amsmath}
\usepackage{amssymb}
\usepackage{graphicx}
\usepackage{subfigure}
\usepackage[inline]{asymptote}
\usepackage[hidelinks]{hyperref}
\usepackage{tikz}
\usepackage{tikz-3dplot}
\usepackage{pgfplots}
\usetikzlibrary{3d, calc}
\usepackage{siunitx}
\makeatletter

\usepackage{color}

\begin{document}

\title{Predictive tracking of the NV Center based on external temperature sensors}
\author{Manpreet Singh Jattana}
\email[Corresponding author: ]{jattana@em.uni-frankfurt.de}
\affiliation{Modular Supercomputing and Quantum Computing,\\Goethe University Frankfurt, Kettenhofweg 139\\ 60325 Frankfurt am Main, Germany}

\author{Thomas Lippert}
\affiliation{Modular Supercomputing and Quantum Computing,\\Goethe University Frankfurt, Kettenhofweg 139\\ 60325 Frankfurt am Main, Germany}
\affiliation{Jülich Supercomputing Centre, Institute for Advanced Simulation, Forschungszentrum Jülich, Wilhelm-Johnen-Straße, 52428 Jülich, Germany}

\begin{abstract} We report an experimental design where the position and resonance frequency of the Nitrogen Vacancy (NV) in a diamond are correlated with the room temperature. A simple model trained on the interpolated correlation data predicts both quantities. The predictive tracking of the NV's location enables continuous operation of the NV quantum computer under ambient conditions for a week without recalibration.  \end{abstract}

\date{\today}

\maketitle 


Over the last two decades, quantum computing has evolved from single-qubit demonstrators to systems with up to hundreds of qubits. There is an intense competition between technologies and research groups have started acquiring small scale hardware.
Some of these competing technologies are cold atoms \cite{bluvstein2022quantum}, superconducting qubits \cite{devoret2013superconducting}, trapped ions \cite{leibfried2003quantum}, quantum dots \cite{burkard, donnelly2024noise}, defects in solids \cite{van2024mapping}, and photonic systems \cite{o2009photonic}. 

We focus our attention on Nitrogen Vacancy (NV) centers. Single defects in diamonds provide a promising platform for quantum computations \cite{DOHERTY20131,276.5321.2012, childress13}. Electron and nuclear spins in an NV count as qubits fulfilling different roles.  While electron spins offer fast control and high fidelity readout \cite{de2010universal,christle2015isolated,sukachev2017silicon,becker2018all}, nuclear spins provide additional qubits with long coherence times that can be used to store and process quantum states \cite{maurer2012room,muhonen2014storing, waldherr2014quantum,CramerNatureComm2016,yang2016high}. Electron qubits play a central role because they can be used to control and connect nuclear spins as well as provide the potential to scale this technology through electron-electron and electron-photon couplings  \cite{yang2016high,togan2010quantum,sipahigil2016integrated,trusheim2018transform,evans2018photon}. The latter can help realize long-range entanglement \cite{bernien2013heralded, humphreys2018deterministic, pompili21}.

Owing to these advantages, NVs have become a reasonable candidate for quantum computations when using optical detection of the magnetic resonance. The sensitivity of NV center's photoluminescence to temperature has been well studied \cite{PhysRevLett.131.086904, PhysRevLett.131.086903}. When operated below $10$~K, spin-dependent optical transitions~\cite{robledo11nature} have demonstrated multipartite entangled states with up to $7$ qubits~\cite{bradley19}. High-fidelity two-qubit gates \cite{dehollain2016bell,van2012decoherence,taminiau2014universal,rong2015experimental, zaiser2016enhancing, unden2018revealing, huang2019experimental}, as well as basic quantum algorithms \cite{van2012decoherence, kalb2017entanglement} have been demonstrated. However, the higher cost and specialization required for cryogenic equipment is a deterrent for those interested in NV quantum computing lacking a strong technical expertise. 

Fortunately, the NV can also easily operate at room temperature, making it very cost competitive for devices with a few qubits. At room temperature, researchers have employed the NV's spin as a sensor for magnetic~\cite{maze08,balasubramanian08} and electric fields~\cite{dolde11}, and thermometry~\cite{kucsko13, acosta10}. 
Operation at room temperature with typical fluctuations of environmental variables like temperature and humidity causes a thermal dislocation of the focus of the optical instruments and the NV, resulting in a photoluminescence loss causing computational errors. Thus, a major obstacle is to periodically determine the NV's position within the diamond with high accuracy. Periodic relocation scans run in three dimensions in turn interrupt quantum computations and reduce computational time.

The problem has conventionally been approached by using a large tracking window (e.g. $5\mu \text{m}\times5\mu\text{m}$ parallel to the surface \cite{PhysRevLett.134.083602}), but this leads to either less accurate position estimates or increased relocation time, without effectively solving the problem. Alternatively, multi-point methods can track with significantly fewer data points. The use of $8-$points \cite{8pointkey, 8key2}, and even $6-, 4-, 3-$point methods has been demonstrated \cite{PhysRevResearch.2.043415, cite9909key, cite55341key}. In nanodiamond thermometry, more points are expected to be better at tracking but at the expense of less measurements which may lead to decline in precision \cite{PhysRevResearch.2.043415}. In quantum computing, any time spent on tracking the NV is lost for actual computation. While the multi-point methods aim to reduce the number of points scanned, they still rely on some form of active relocation measurements, interrupting any form of computation. Overall, despite a reduction in the track-and-locate time using multi-points, these methods essentially remain in the paradigm of stop-and-fix, where the computation is stopped to fix the tracked coordinates.

In this work, we determine the NV's position without requiring active scanning, allowing for continuous operation. We report on the mechanism of placing the optical equipment in thermal isolation leading to the correlation of the NV's position and resonance frequency with the room's temperature sensors. By training a simple model from the experimental data, we are able to accurately predict the positions and resonance frequency $\nu_\mathrm{res}$ using external temperature measurements as inputs. As we show below, this design and implementation is a simple yet highly effective way to reduce the time spent to periodically track the electron within the NV at room temperature inside a tight window ($1\mu\text{m}\times0.7\mu\text{m}$) parallel to the surface. Our work moves the recalibration into a different paradigm: a simple experimental redesign that enables modelling the drift and predicting it in real-time. The paradigm is moved from stop-and-fix to predict-and-prevent. The location of the NV is predictively tracked and the necessity for scanning is prevented for prolonged periods of time.

\begin{figure}[h]
	\centering
	\begin{tikzpicture}
		\node at (-0.15,-0.17) {\includegraphics[width=0.45\textwidth]{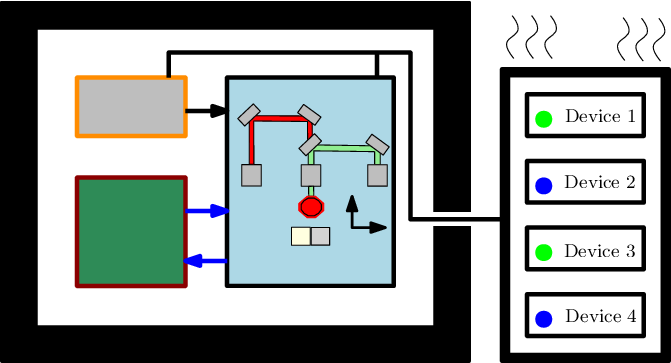}};
		\node at (-0.4,0.9) {Optics table};
		\node at (0.4,-0.9) {x};
		\node at (-1.05,-0.8) {\tiny{Magnet}};
		\node at (-1.15,-0.1) {\tiny{D}};
		\node at (-0.45,-0.1) {\tiny{O}};
		\node at (0.35,-0.1) {\tiny{B}};
		\node at (0.18,-0.35) {z};
		\node at (-2.55,0.75) {Laser};
		\node at (-1.55,-1.7) {\footnotesize{T$_2$ (room)}};
		\node at (-0.5,-1.2) {\footnotesize{T$_1$ (table)}};
		\node at (2.85,1.45) {Heat};
		\node at (-2.58,-0.55) {Temp.};
		\node at (-2.58,-0.85) {control};
		\node[text=white!80] at (-1.4,1.85) {Weak thermal isolation};
	\end{tikzpicture}
\caption{Experimental optical setup of our NV center based quantum computer. The optics table, laser, and liquid cooling temperature control are placed within a weakly thermally isolated room. The optics table consists of a diamond (red) placed close to a magnet. A green beam of light originating at \textbf{B} is shone onto the diamond through the objective \textbf{O}. Emitted red light is detected at detector \textbf{D}. Control equipment consisting of several devices that generate most of the heat are kept outside.}
	\label{fig:eps_example}
\end{figure}
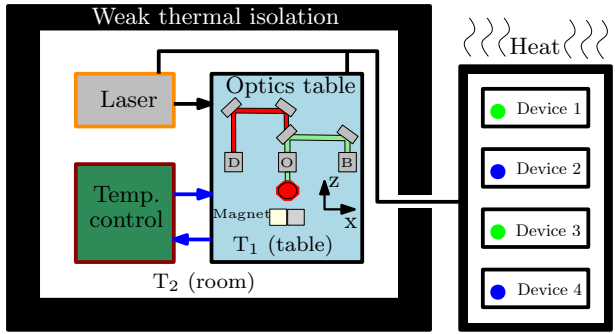

The experimental setup is designed such that the optical table is enclosed in a box of dimensions $45$ cm $\times~50$ cm$\times~15$ cm (L x W x H). Figure~\ref{fig:eps_example} shows the setup. The optics table is located in a small room of dimensions around $2$ m~$\times~2$ m$~\times 3$ m that has a weak thermal isolation, i.e. no use of heat exchange equipment. 
The optics table is liquid cooled and kept at a constant target temperature. A sensor is placed on the optics table to measure the temperature $\text{T}_1$. 
The major sources of heat generation, e.g. pulse generators, are kept outside in a bigger room which is air cooled. T$_2$ is the room temperature measured through a single sensor mounted in the room and varies uncontrolled with the temperature outside. Both $\text{T}_1$ and $\text{T}_2$ are not to be confused with spin relaxation or coherence times and are measured every $20$ seconds.

Our NV of interest is located within a few microns of the surface of an ultrapure diamond sample. The electron spin has a gyromagnetic ratio of $28.024$ GHz/Tesla and has a zero field splitting at $\approx 2.869$ GHz. The external magnetic field is fixed at $\approx 0.05$ Tesla. We track the NV in a window of dimensions $1 \mu\text{m}, 0.7 \mu\text{m}, \text{and } 3 \mu\text{m}$ in the $X,$ $Y ,\text{ and }Z$ orientations, respectively, every $400$ seconds using the software Qudi \cite{BINDER201785}.

\begin{figure*}
\centering
\begin{tikzpicture}
\node[anchor=north west] (imgA) at (0,0)  {\includegraphics[scale=0.58,trim={0.5cm 0cm 0cm 0cm}, clip]{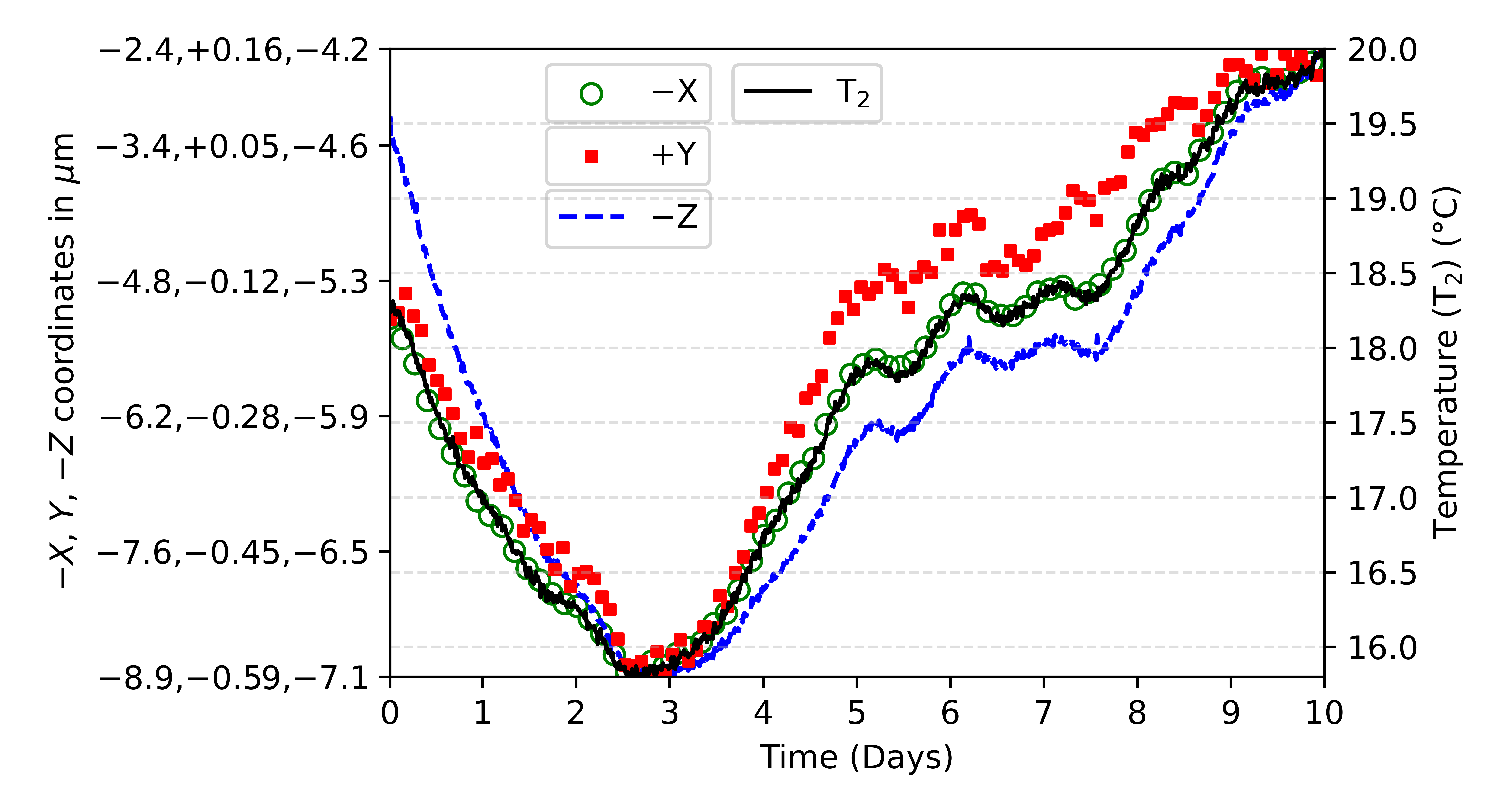}};
\node[anchor=north west] at ([xshift=60pt,yshift=5pt]imgA.north west) {\textbf{(a)}};

\node[anchor=north west] (imgB) at (14.6,-0.2)
{\includegraphics[scale=0.5, height=4.8cm]{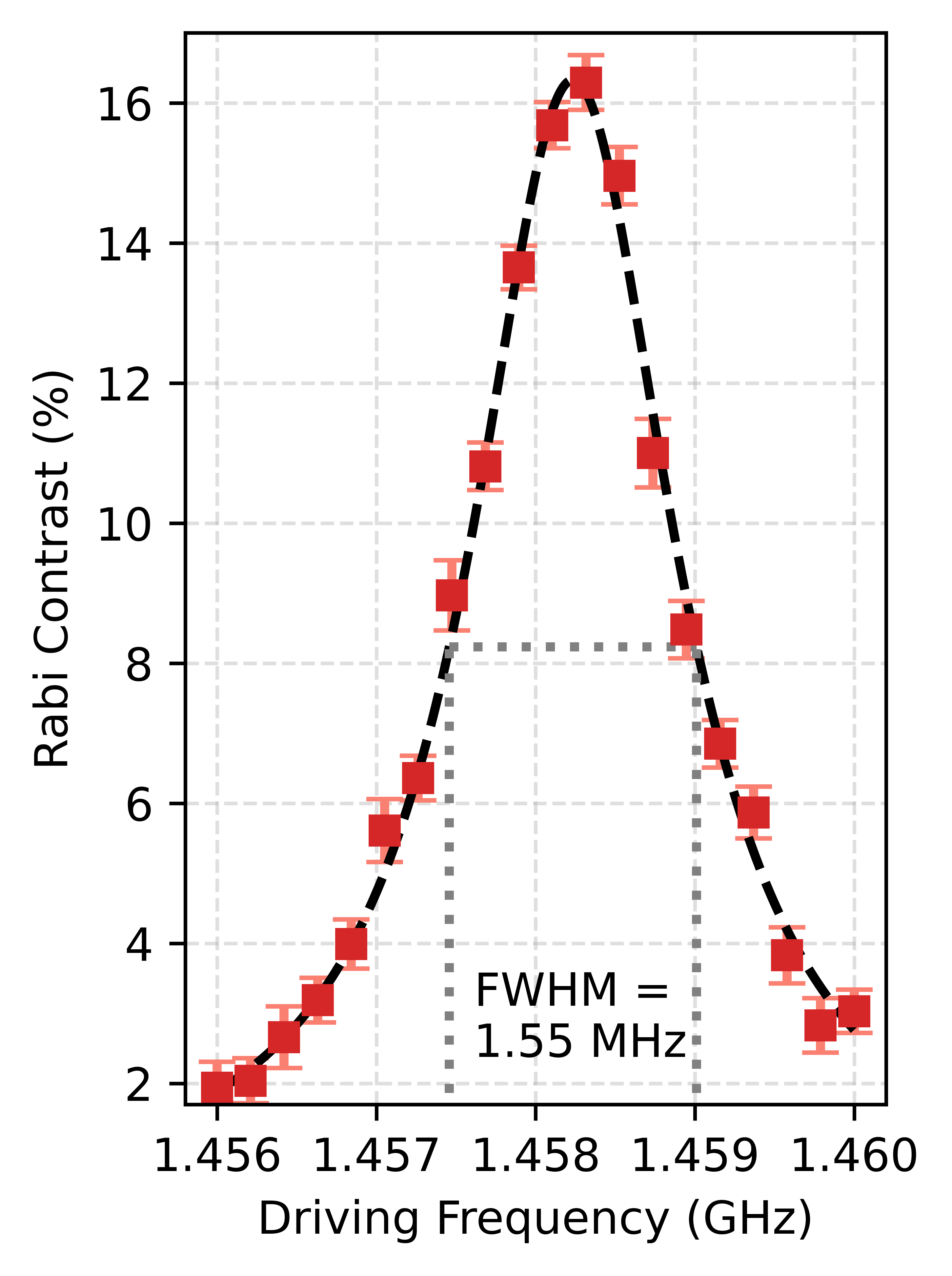}};
\node[anchor=north west] at ([xshift=18pt,yshift=10pt]imgB.north west) {\textbf{(c)}};

\node[anchor=north west] (imgC) at (9.0,-0.3) {\includegraphics[scale=0.28, trim={0.0cm 0cm 0.4cm 0.0cm}, clip]{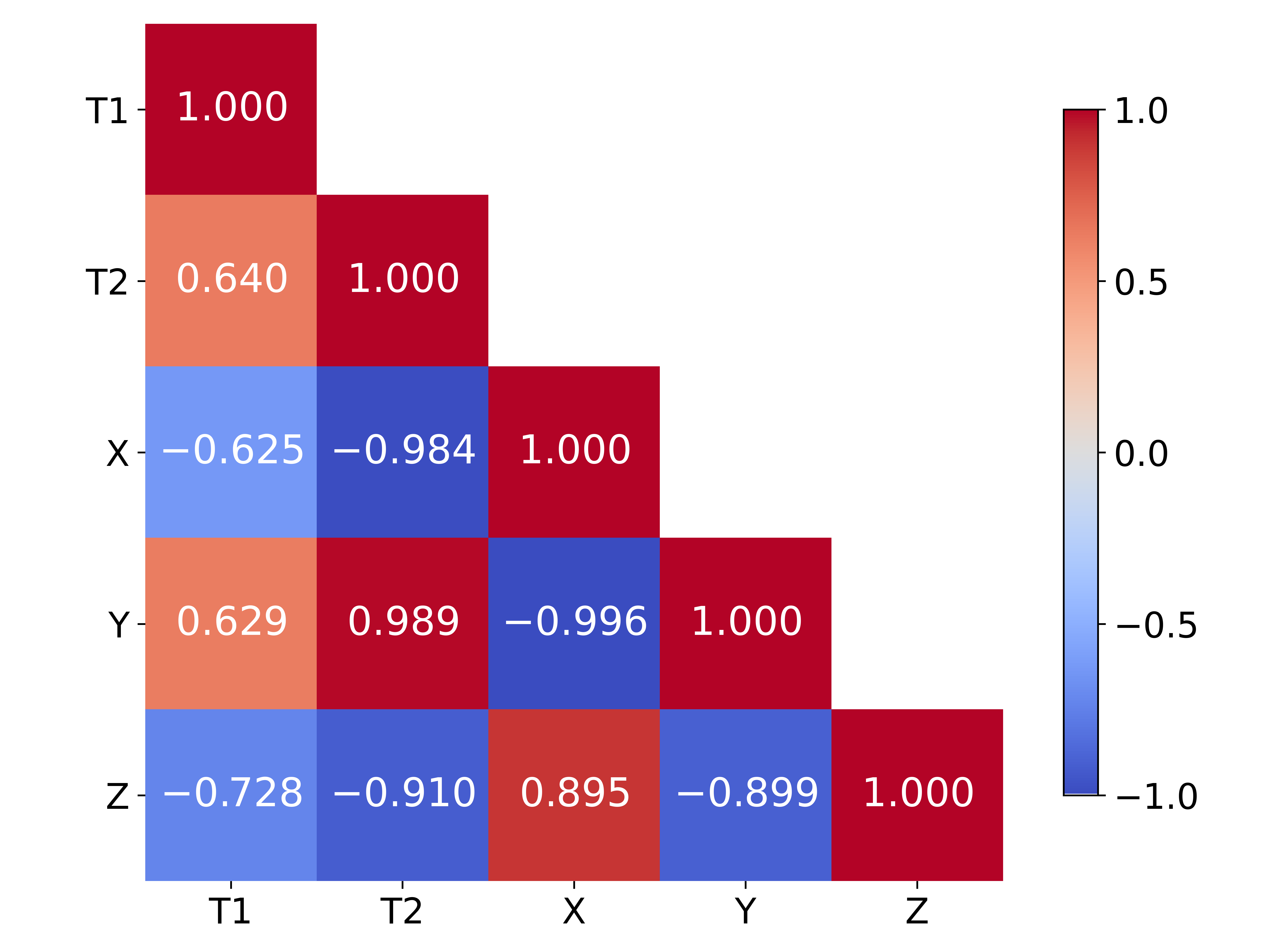}};
\node[anchor=north west] at (9.7,0.15) {\textbf{(b)}};
\end{tikzpicture}
\caption{\textbf{(a)} The relative change in NV's location (left axis) and temperature T$_2$ (right axis) plotted against a time period of $10$ days. The values in $X$ and $Z$ directions are negated for a better visualisation of the anticorrelation. There are strong linear (anti/)correlations despite the different scales of $X$, $Y$, and $Z$. The axes are not optimized for superimposition. \textbf{(b)} The correlation matrix between the NV's coordinates $X$, $Y$, and $Z$, and temperatures T$_1$ and T$_2$ shows the strength of their linear correlations (red) and anticorrelations (blue). The apparent stronger (anti/)correlation of positions with T$_2$ compared to T$_1$ is due to the measurement sensitivities of the sensors.
As is also visible in (a), the measurables $-X$, $Y$, and T$_2$ have near perfect correlation. \textbf{(c)} The percentage difference in the crest and trough of the Rabi oscillations obtained for different driving frequencies at constant temperature is plotted in red square points. 
A Lorentzian fit to the data is shown in dashed black lines. The full width at half maxima (FWHM) is $1.55$ MHz.  \label{fig:combined}}
\end{figure*}
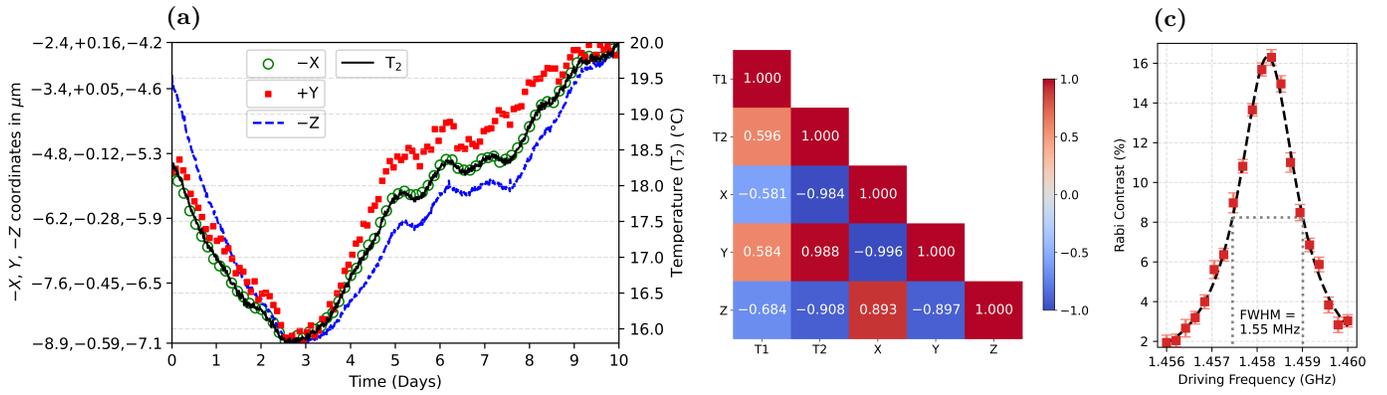

We continuously track the change in the location of the NV w.r.t. the temperature over a period of $10$ days. The lowest and highest temperatures measured were $13.08$°C and $13.31$°C for T$_1$, and $15.8$°C and $20.0$°C  for T$_2$, respectively. It is worth noting that T$_1$ and T$_2$ have different least counts, namely, $0.01$°C and $0.02$°C, respectively. The point of origin of the NV's coordinates is chosen arbitrarily.
Data collection was interrupted for ten minutes on day three by an unexpected system restart. This caused an unusually large change in $Y$ and $Z$, attributed to an unknown software glitch during the abnormal shutdown.
For data normalisation, we added $-0.227\mu$m and $0.05\mu$m to all $Y$ and $Z$ values after the third day, respectively, such that they had identical positions in the last pre- and first post-interruption measurements.

We plot the change in the NV's location w.r.t. T$_2$ in Fig.~\ref{fig:combined}(a). The left y-axis gives the positions in $-X$, $Y$, and $-Z$ coordinates, respectively, in units of $10^{-6}$ m. The values of $X$ and $Z$ are negated in order to better visualize their anticorrelation w.r.t. T$_2$. We perform a linear interpolation of all the four variables since the temperature data and NV's location is not collected at exactly the same time but often with a difference of less than few tens of seconds. Fig.~\ref{fig:combined}(a) shows only a finite sample of these interpolated data points for visualization reasons. 

To better understand the relationship between the NV's location and the temperatures, we linearly interpolate the data points to calculate any correlations between the variables. First, we normalise the data of all $5$ variables such that it ranges between $0$ and $1$. Second, we rearrange T$_2$ values in the decreasing order and sort all the other variables according to this order. Doing so allows us to retain the same timestamped values of others variables in reference to T$_2$. Third, we do a linear fit to all the variables to compute the respective slopes. Fourth, we use $1 - \frac{ 2\left| \, |a| - |b| \, \right| }{ \left( |a| + |b| \right) }$ or $ \frac{ 2\left| \, |a| - |b| \, \right| }{ \left( |a| + |b| \right) } -  1$ when slopes $a$ and $b$ have the same signs or opposite signs, respectively, to compute the correlations. Finally, we calculate these correlations for all pairs of variables and show the results in Fig.~\ref{fig:combined}(b).

Data that shows the same trend will have the same slope and hence will be perfectly correlated through our formula. From Fig.~\ref{fig:combined}(b), we observe that the positions $X$ and $Z$ ($Y$) are exceedingly well anti-correlated (correlated) with the temperature T$_2$, and only weakly with T$_1$. The weak correlation is likely an artifact of the small temperature change (0.23~°C) subjected to a limited dynamic range because of the sensor's least count (0.02~°C). Fig.~\ref{fig:combined}(b) quantifies the visual relations shown in Fig.~\ref{fig:combined}(a). Furthermore, the positions $X$, $Y$, and $Z$ are strongly (anti-)correlated among themselves, as is to be expected.

\begin{figure}
	\includegraphics[scale=0.57]{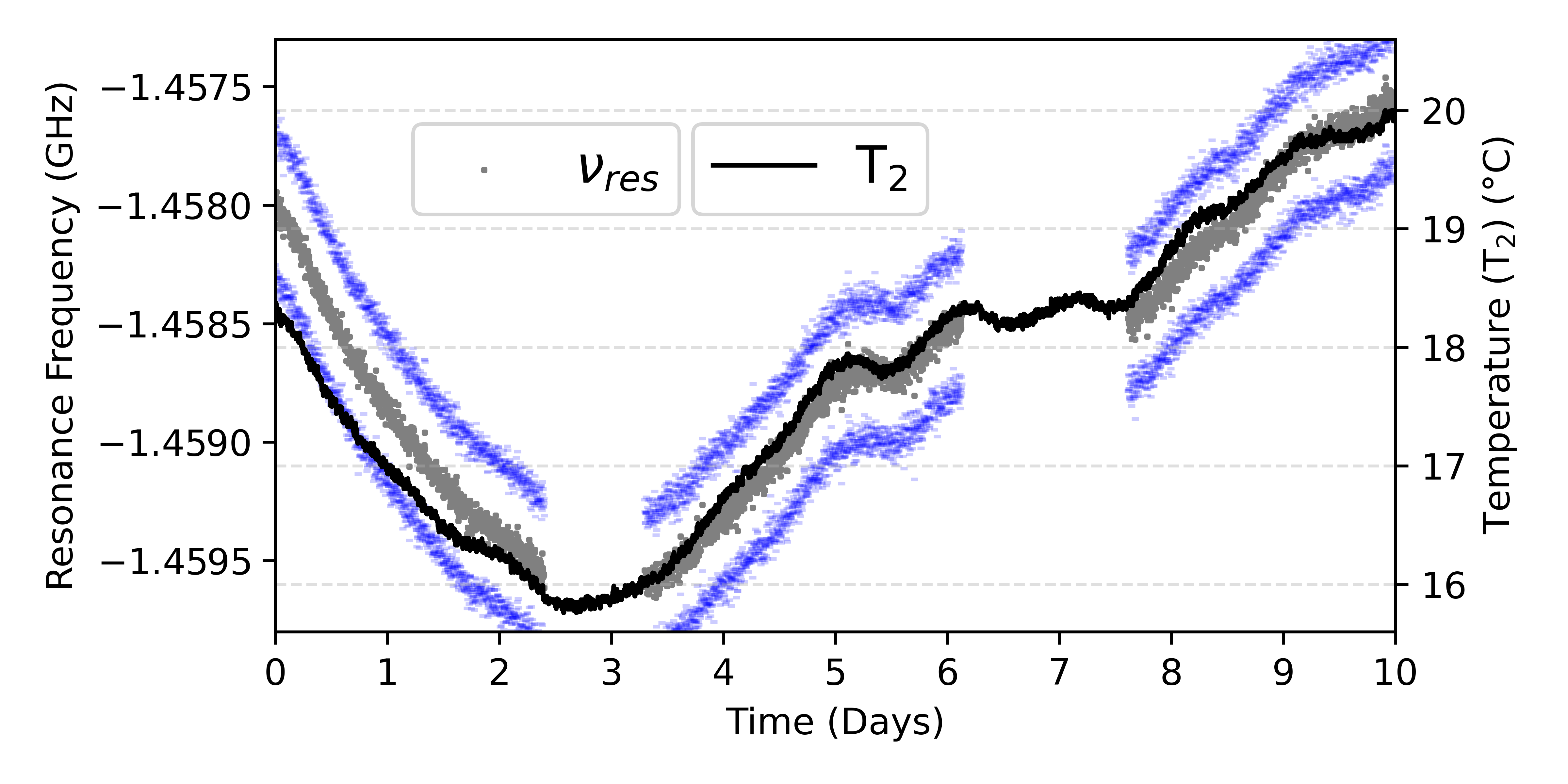}\caption{The negative of measured resonance frequencies (grey points) in GHz (left axis) and temperature in °C (black line) T$_2$ (right axis) are plotted against time.  Each blue point represents one standard deviation of the resonance frequencies.  \label{odmr}}
\end{figure}

\begin{figure}
\begin{tikzpicture}
\node[anchor=south west, inner sep=0,  xshift=4pt] (image) at (0.01,0) {%
\includegraphics[scale=0.6]{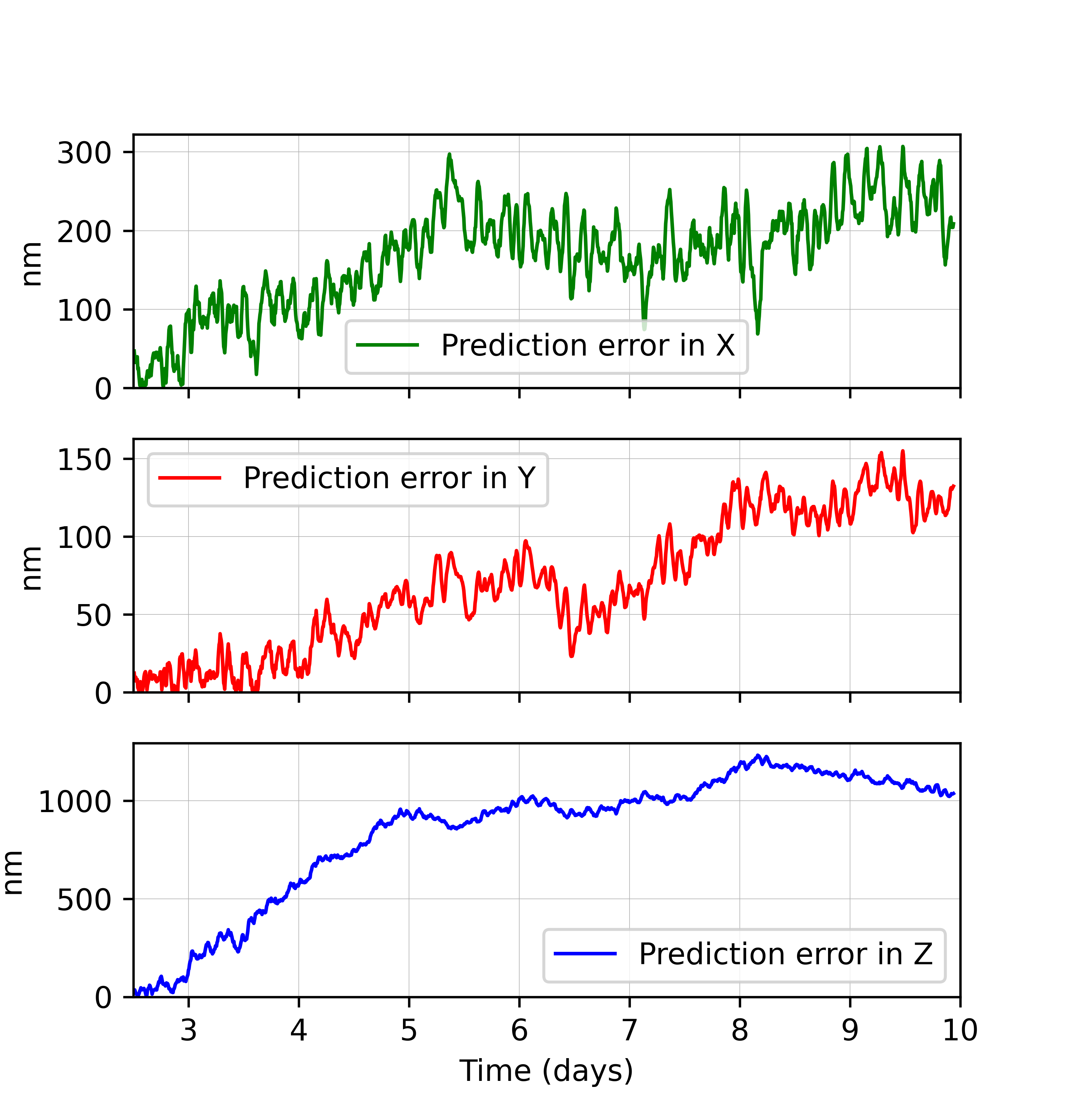}%
};
\node[anchor=north west, inner sep=0] (image_bottom) at ([yshift=5pt,xshift=1pt]image.south west) {%
    \includegraphics[width=7cm, height=3.2cm]{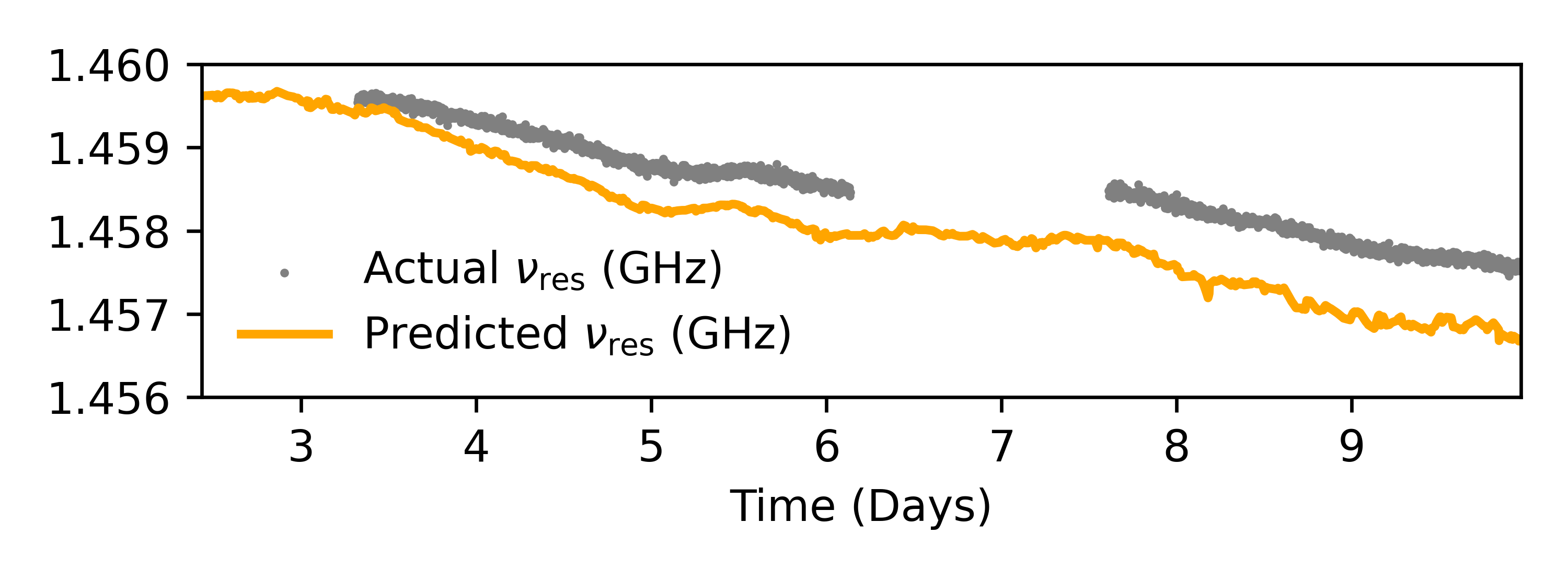}%

};
\begin{scope}[
x={(image.south east)},
y={(image.north west)}
]
\draw[black, thick] (0.01, -0.35) rectangle (0.95, 0.950);
\draw[black, thick] (0.005, 0.01) rectangle (0.975, -0.35);
\node[anchor=north west] at ([xshift=0pt,yshift=215pt]imgA.north west) {\textbf{(a)}};
\node[anchor=north west] at ([xshift=0pt,yshift=-64pt]imgA.north west) {\textbf{(b)}};
\end{scope}
\end{tikzpicture}
\caption{\textbf{(a)} The absolute difference between actual and predicted $X$ (top), $Y$ (middle), and $Z$ (bottom). The units of y-axes is nanometers.  \textbf{(b)} The actual $\nu_\mathrm{res}$ (grey) in GHz and the predicted $\nu_\mathrm{res}$ (orange) as a function of time. \label{predict}}
\end{figure}

We now investigate the question: 
to what extent are the Rabi oscillations of the electron sensitive to the driving frequency? We calculate the \textit{Rabi contrast} as two times the amplitude of a sine fitting to the oscillations divided by the offset. We run a scan between $1.456$ GHz and $1.460$ GHz and plot the percentage Rabi contrast in Fig.~\ref{fig:combined}(c). The data points fit very well to a Lorentzian curve and the full width at half maxima is $1.55$ MHz. This measure gives us a quantitative loss in Rabi contrast (detuning error) one is willing to tolerate for off-resonance frequencies.

If the NV's location changes due to a temperature change, what is the change in the resonance frequency? We continuously measure the resonance frequency for a large portion of the $10$ days with only two interruptions. Each measurement takes $\approx 191$ seconds. The results are shown in Fig.~\ref{odmr}. The measured resonance frequencies (in GHz) along with their standard deviations and temperature T$_2$ are plotted against time. We visually observe a strong first-order linear correlation between T$_2$ and the resonance frequency.


We use second order polynomial regression to predict the position coordinates of the NV and the resonance frequency $\nu_\mathrm{res}$. We train our models on the data T$_1$ and T$_2$ suitably linearly interpolated across the first two and half days so that the data points have same timestamps. 
We use the model
\[
\alpha_{\gamma} = \beta_{0,\gamma} + \beta_{1,\gamma} \text{T}_1 + \beta_{2,\gamma} \text{T}_2 + \beta_{3, \gamma} \text{T}_1^2 + \beta_{4,\gamma}\,\text{T}_1\text{T}_2 
    + \beta_{5,\gamma} \text{T}_2^2,
\]
where we estimate the coefficients $\beta_{i,\gamma}$ by minimizing least squares for each ${\gamma}\in \{X,Y,Z,\nu_\mathrm{res}\}$ separately. The choice of the model is motivated by the results shown in Fig.~\ref{fig:combined}(a) where some non-linear contributions are apparent to mostly linear correlations. With ordinary least squares, these parameters are optimized to best fit the interpolated observed data and capture both linear, using $\beta_{0,\gamma}, \beta_{1,\gamma}, \beta_{2,\gamma}$, and non-linear, using $\beta_{3,\gamma}, \beta_{4,\gamma}, \beta_{5,\gamma}$, contributions. The goal of the modeling is to accurately and precisely predict $\nu_\mathrm{res}$ and coordinates $X$, $Y$, and $Z$ given T$_1$ and T$_2$. The accuracy and precision of prediction for the coordinates is related to the size of the tracking window in which the NV's position is scanned. 

The results for the coordinate predictions are shown in Fig.~\ref{predict}(a). The plots show the absolute difference between the actual data and the prediction data of the model. The results show that our model's accuracy is different in each coordinate. The predictions become less accurate over time, which is to be expected. While the model is able to predict locations for coordinates $X$ and $Y$ rather accurately, it only qualitatively captures the movement of $Z$. The underlying cause requires further investigation. However, the prediction error in $Z$ is still well within the size of the tracking window.

In Fig.~\ref{predict}(b), we show the prediction for $\nu_\mathrm{res}$ along with the actual values. 
We observe that our model is able to predict the change in $\nu_\mathrm{res}$ within about one FWHM (see Fig.~\ref{fig:combined}(c)) of Rabi contrast for slightly more than a week when temperature is left uncontrolled. The cause of the frequency drift is most likely the change of distance between the magnet and the diamond due to the thermal expansion/contraction of the optics table on which they are mounted. We emphasize that for continued normal operations of an NV center based quantum computer, it is only required that $\nu_\mathrm{res}$ is predicted accurately within a few hours, not days. Our simple model is sufficient for this purpose.

Albeit highly setup-specific, it is interesting to calculate the shift in NV's position per °C by dividing the maximum change in each coordinate by the maximum change in T$_2$ across the entire data. The position changes by $\Delta X = 1.533 \mu \text{m/°C}, \Delta Y = 0.178 \mu \text{m/°C}$, and $ \Delta Z = 0.682 \mu \text{m/°C}$. Similarly, $\Delta \nu_\mathrm{res}$ = $517$ kHz/°C. Thus, for a temperature change of $>0.33$ °C, $>1.98$ °C, and $>2.21$ °C, the brightest spot of the NV's point spread function (PSF) will be outside our tracking window in the $X$, $Y$, and $Z$ directions, respectively. Similarly, a temperature change of $1.41$ °C reduces the Rabi contrast by half. These values give us a reference for our experiments to which, if the external temperature T$_2$ is optimally controlled, and everything else remains unchanged, the NV will always be found within the tracking window and reachable at the same resonance frequency with a high contrast.

If T$_2$ varies uncontrolled, as in our experiment, the trained model allows one to predict the NV's movement without requiring periodic active position measurements. The NV's position can then be updated, without active scanning, at regular intervals by the model's predictions based on T$_1$ and T$_2$. This prevents the interruptions of quantum algorithms and allows for continuous operation of the quantum computer. While our model takes into account the temperature fluctuations and makes reasonable predictions for the position coordinates, temperature is not a unique factor influencing the NV's position. Factors like humidity, diamond strain, chiller flow rate, and table vibrations should be investigated and potentially added to the simple model for more precise predictions.   

Besides quantum computing, our method can be useful in nanodiamond thermometry where multi-points methods are currently used. If the nanodiamonds are immobilized, for instance by surface adsorption, polymer embedding, or covalent attachment to a substrate, their Brownian motion can be effectively suppressed, and our approach could in principle be applied in this scenario for real-time estimations of magnetic resonance shifts in biological applications.

We described a simple experimental setup where an NV center in a diamond is placed in weak thermal isolation correlated very well with temperature sensors for measurements taken over ten days. We demonstrated that the resonance frequency is also correlated to the room temperature. By using a simple quadratic model as a function of the temperature of two sensors, training it for a period of two and half days, we were able to accurately and precisely predict the movement of the NV for the next one week. Similarly, our model was able to predict $\nu_\mathrm{res}$ to a reasonable accuracy.

Our results can be improved in several ways. More temperature sensors with lower least counts and shorter logging intervals would help generate more accurate data for training our model leading to more accurate predictions. Different sensors, e.g. to measure humidity and vibrations, could also be integrated in the model. The model itself could be made more complicated to include non-linear correlations more accurately.

We thank Gopalakrishnan Balasubramanian, Priya Balasubramanian, Matthias Gerster, and Florian Frank from XeedQ GmbH for their generous support around the hardware setup. We also thank Philip Döbler for his feedback.

\bibliography{reference}

\end{document}